# Towards a potential paradigm shift in health data collection and analysis

## *Contemporary challenges of human-machine interaction*

*David Josef Herzog*
*Computer Science,*
*London campus of*
*Ulster, Solent, and Roehampton Universities,*
*London, UK*
*Davidjosef.herzog@qa.com*

*Nitsa Judith Herzog*
*Data Science Technology*
*Faculty of Engineering and Environment*
*London campus, Northumbria University,*
*London, UK*
*Nitsa.herzog@northumbria.ac.uk*

**Abstract**

Industrial Revolution 4.0 transforms healthcare systems. The first three technological revolutions changed the relationship between human and machine interaction due to the exponential growth of machine numbers. The fourth revolution put humans into a situation where heterogeneous data is produced with unmatched quantity and quality not only by traditional methods, enforced by digitization, but also by ubiquitous computing, machine-to-machine interactions and smart environment. The modern cyber-physical space underlines the role of the person in the expanding context of computerization and big data processing. In healthcare, where data collection and analysis particularly depend on human efforts, the disruptive nature of these developments is evident. Adaptation to this process requires deep scrutiny of the trends and recognition of future medical data technologies` evolution. Significant difficulties arise from discrepancies in requirements by healthcare, administrative and technology stakeholders. Black box and grey box decisions made in medical imaging and diagnostic Decision Support Software are often not transparent enough for the professional, social and medico-legal requirements. While Explainable AI proposes a partial solution for AI applications in medicine, the approach has to be wider and multiplex. LLM potential and limitations are also discussed. This paper lists the most significant issues in these topics and describes possible solutions.

## 1. Introduction

World demographic trends require adjustment of all social institutions. The rapid population growth in expanding economies coexists with lower birth rates and quick ageing in economically developed societies [1]. Healthcare systems have to adapt to the higher demand. Extensive broadening of healthcare services cannot be seen as a primary solution. The fourth technological revolution is projected to be one way to solve the problem of society's growing financial and social healthcare burden [2, 3].
New technologies have improved communication, data collection, and management. Artificial intelligence helps analyze voluminous medical data, ubiquitous computing, and Ambient Assisted Living to improve the condition of home-based patients. Telemedicine and robotics connect geographically distant specialists and patients, to name just a few. The SARS-CoV-2 pandemic inadvertently intensified current trends [4].

At the same time, the technological development of healthcare has to overcome not only financial, social and psychological barriers. There are fundamental and methodological problems which cannot be solved by financial and administrative efforts. Human-machine interaction specialists are acutely aware of the significant gap in effective synergy [5]. Big data studies are fruitful in medical image processing, laboratory analysis, public health, and some areas of clinical research, but they are somehow less fruitful in the field of general patient-centred clinical healthcare data analysis. The pandemic demonstrated the importance and robustness of prophylaxis on the level of public health by means of quarantine and vaccination. On the personal level, besides prevention, treatment is exhaustively thorough in intensive care, with traditional oxygen, heavy staff workload, and pulmonary and other intensive care protocols of treatment. The highly effective drug for direct treatment has been developed only recently, despite all endeavours with the help of the best technologies [6].

To understand the underlying causes, it is important to disentangle fundamental issues of human-centred technologies from temporary "development" hindrances [5]. Appropriate focus on the right type of solutions allows for proper healthcare progress. Figure 1, provided below, demonstrates the structure of the represented topics.

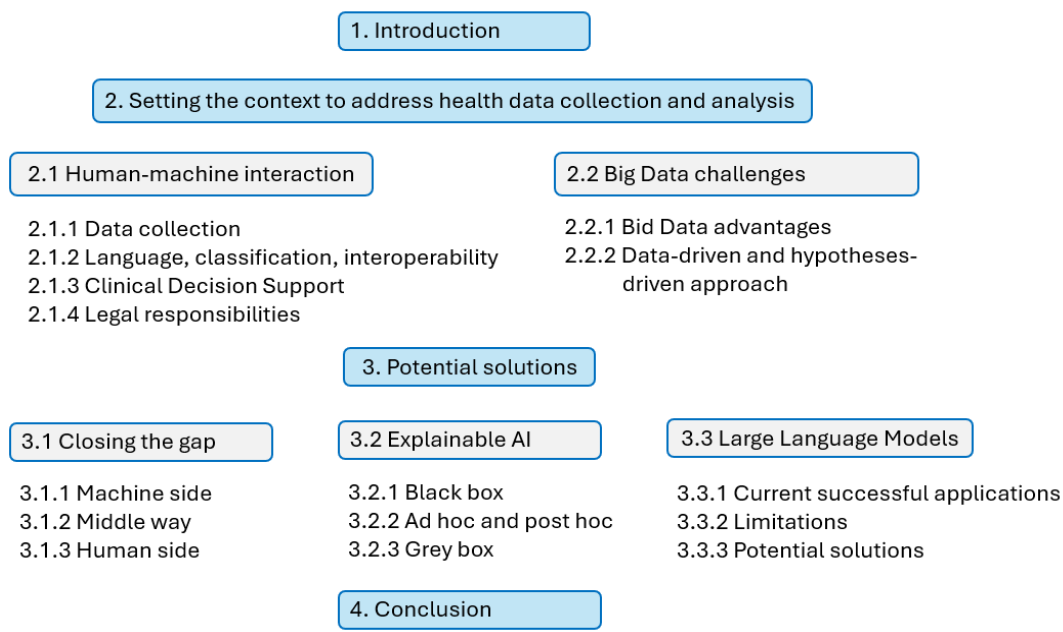

*Figure 1*: The article's roadmap

## 2. Setting the context to address health data collection and analysis

There are a number of fundamental and applied questions, which are understandably demanding focus before any solution can be found. Human-machine interaction is reshaped by ubiquitous computing and big data. The main areas in these topics hold keys to the balanced vision and rational estimation of future changes in the area of healthcare data handling.

### 2.1 Human-machine interaction

HMI was formulated in the 1960[th] and can be distilled into the necessity to interact between user and machine despite their dissimilarities [7]. There are several areas in the HMI: language disparities between human and technical devices, different types of possible errors and bias, and demand for specific interfaces, humans and machines, to list just a few. Ideally, technologies would allow the closing of the gap between humans and machines by creating perfect interactive systems, but this task has yet to be fulfilled.

## 2.1.1 Data collection

Historically, information was collected, recorded, and analyzed by humans only, first at the biological and psychological levels and then at the social and civilization levels [7]. Language, writing, mathematical and statistical methods allowed data prepossessing, classification and analysis. The advance of machines and information theory laid the foundation for the computer era. HMI problem was formulated.

From the start, data collection was prone to human mistakes and bias. Input into the machine was highly technical and required an appropriate interface, which could create additional obstacles for human operators. Then, regardless of the form of data storage, the problem of human readability arises. Health and biometric information are mostly collected by humans, and the records are still mostly human-readable. Machine intervention requires significant effort to optimize data collection as a prerequisite for machine storage and analysis. The overbearing position of humans is inevitably being taken more and more by machines with improvements in data collection, archiving, transmission, and analysis. Human control is important in clinical data handling, but technical assistance is clearly beneficial [8].

Today, healthcare data is potentially heterogeneous and comes not only from primary health records. Secondary data is harvested from Electronic Health Records (EHR). The three basic Vs of big data, volume, velocity, and variety, require a different approach. Behavioural data flows from the Internet and surveillance sources, and more information comes from smartphones, smart health devices and applications, smart homes, and smart environments. Clinical data itself, especially biometric data, is often read from sensors before being recorded by medical staff in EHR [9].

## 2.1.2 Language, classification, interoperability

There is a well-known fundamental difference between human natural language and machine languages. Automatic data analysis also employs classification techniques. Artificial Intelligence classification of health data significantly differs from the traditional human-based one. Human bias is reflected, for example, in nosology's diseases` classifications. Polysemy in terminology creates confusion even in the different areas of medicine. An unclear definition of the simple term "abdomen" prevents the easy transfer of information [10]. Other healthcare areas, like psychiatry, are known for difficulties in precise classification [11].

Medical nosology classifications historically reflect expert-driven beginnings, objective observations, and further classifications based on symptoms and syndromes. Anatomy, physiology, biometric information, and genetic and biomolecular data today form the basis for more acceptable bottom-up classifications [12]. However, being ontologically more modern, these classifications still lack easy machine readability. Figure 1 schematically shows the standard medical procedure from obtaining initial information to the final records. From this process, we can see that nosology classifications naturally depend on expert-driven processes.

While the initial trigger can be complaints by patients, relatives or members of the public, the next step is always framed by medical professionals, even when the starting point is a routine examination. After collecting anamnesis, physical, instrumental, and laboratory examinations are performed. The results lead to a primary, most probable diagnosis. They are registered in medical records. Differential diagnosis may require additional checks. Primary nosology code is added to the records. The treatment can be started at the earliest possible stage or postponed to a later stage. The results of treatment also help to make a more precise diagnosis and final detailed diagnosis. A nosology code is applied for final records and stored in the documentation and database.

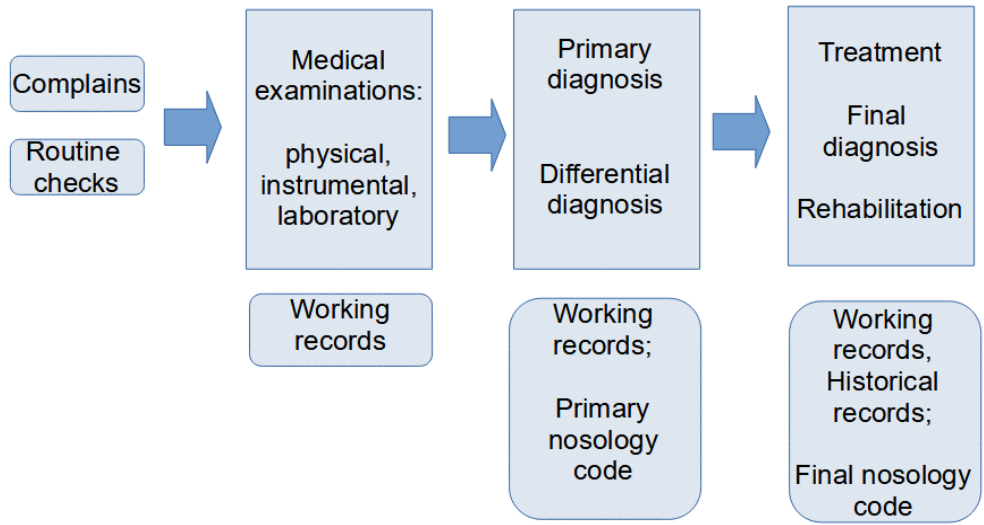

*Figure 2: Standard procedure of diagnosis and treatment, with matching records*

Even top standards differ depending on their classification purpose. The widely used SNOMED-CT terminology is multiaxial, clinically oriented, procedure-oriented, and input-based. The International Classification of Diseases (ICD) is statistically and output-oriented. ICD-10 partially differs in ontology from ICD-11—the last one is closer to SNOMED and multiaxial. However, it does not fully solve the problem of interoperability [13].

Figure 2 shows simplified data flow in healthcare. The starting point is similar to the scheme in Figure 1. Data flow between medical stakeholders` domains is demonstrated by dark blue arrows, while intra-domain data transformation interchange is shown by thin arrows. Light blue blocks reflect generative and transformative nodes. Dark blue blocks show data storage and types of data. We can see complex quasi-horizontal interaction between domains of the general public, healthcare professionals, and biomedical and clinical researchers. The research domain is important for creating treatment standards and protocols, which reflect combined results of analyzed data from the healthcare system and scientific exploratory quasi-horizontal interaction between domains of the general public, healthcare professionals, and the field. Natural language-processed records might be relatively easy to save on the primary level but have to be integrated into machine-readable ontologies on the database level. Analytical results from the databases also cannot be fully machine-adapted because of the inclusion of natural language. The interplay between machine and human readability will exist well in the healthcare system and scientific exploratory quasi-horizontal interaction between domains of the general public, healthcare professionals, and the future.

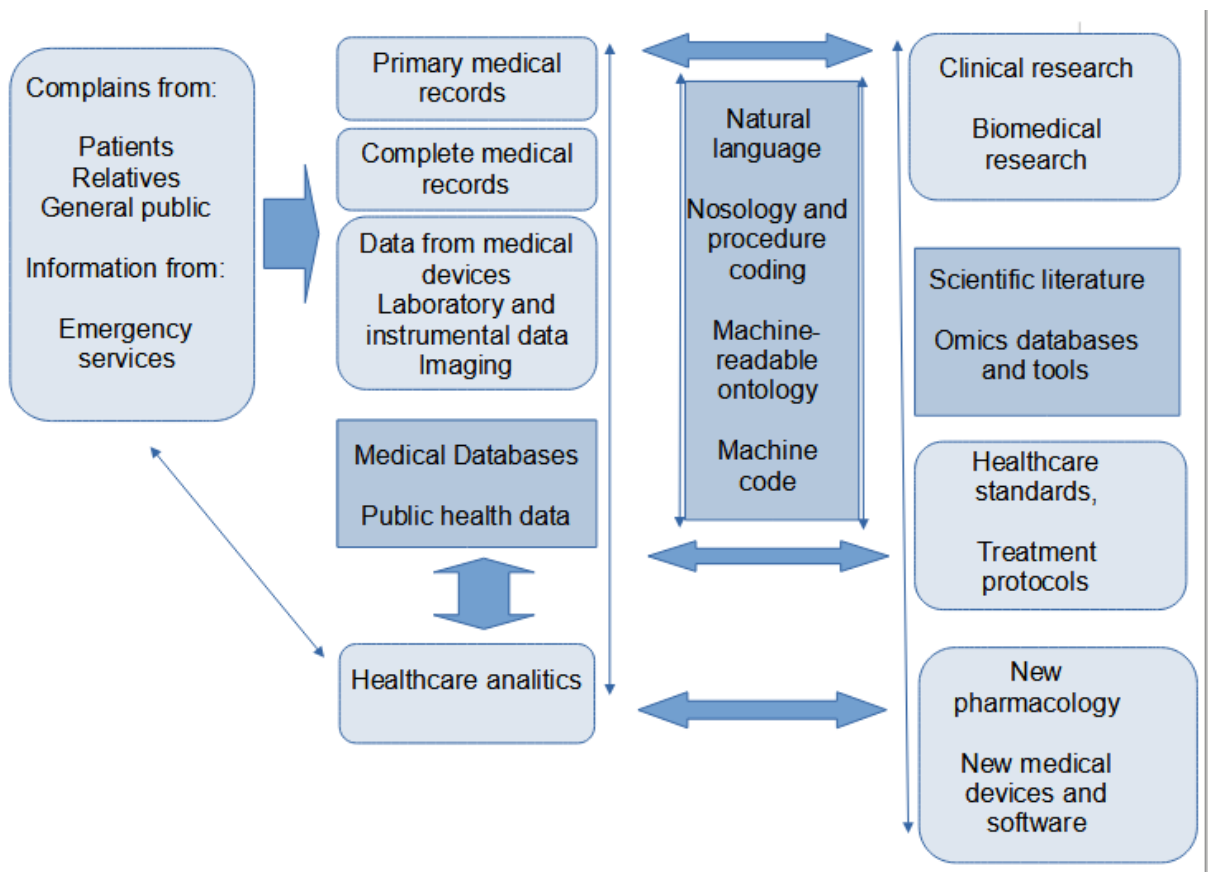

*Figure 3: Data flow and the problem of interoperability*

**2.1.3 Clinical decision support**

Clinical Decision Support System is a software application that helps the main stakeholders make optimal decisions. Patients, administrative staff, and healthcare practitioners – each group sees their own benefits in it [14]. For clinicians, the main purpose of CDSS is to optimize the diagnostic procedure. A widely discussed topic is the ability of the automatic system to diagnose and propose treatment without human supervision. However, Clinical Decision Support Systems (CDSS) are widely available. While expert society is aware of its limitations, there is a gap in understanding when we take the viewpoints of different stakeholders. A significant discrepancy exists between technical and healthcare stakeholders.

Technically, there is no obstacle to maximally automatizing the procedure. The patient can contact CDSS before first interacting with clinical staff. If the data stream from the smart environment is informative enough, a system that can point out possible pathology signs and recommend the next step is conceivable.

This approach is more proactive than the contemporary retroactive point-of-care procedure. The process usually starts with subjective complaints by patients or relatives and goes through iterative stages. Knowledge-based CDSS gives significant advantages at every step. However, it is still too early to speak about a more automatic, proactive, highly autonomous system, even with the help of Big Data and Artificial Intelligence.

**2.1.4 Legal responsibility**

Clinical data usually requires approval from several stakeholders for handling. In the case of private entrepreneurship, rules can be stricter or more relaxed, depending on the position of authorities and society. With the entry of Big Data and Artificial Intelligence, the question of responsibility becomes more complex [15].

It is necessary to understand every step in the data handling process clearly. The nature of Big Data itself makes it more complicated. It is important to point out that artificial intelligence tools often work with unlabeled or partially labelled data when analytical processes are at least foggy or treated as a black box. The legal or technical solutions for these problems are not simple.

### 2.2 Big Data challenges

Big Data is changing society's landscape. Some expect it to replace earlier data collection and analysis techniques. It is difficult to deny the advantages of Big Data and new methods of its analysis. On the other hand, there are reservations about the paramount nature of analytical methods based on Artificial Intelligence. It is possible that some areas of life are more suitable for this approach and some less.

#### 2.2.1 Big data advantages

Big Data, with its basic three Vs—volume, velocity, and variety—has many benefits over traditional forms of data. The fourth V, validity, eliminates the negative influence of bias and brings it to a high-quality level [16]. The speed of data influx allows quick response, which is important in emergency situations and can literally save lives in healthcare.

In public medicine, Big Data may show tendencies of epidemiological processes and trigger an administrative reaction. Biomedical research gains a lot from hypothesis formulation based on big data analysis. Genomics, proteomics, and metabolomics routinely employ Artificial Intelligence for significant masses of data. Behavioural data obtained from the internet, analysis of search engine requests, smart vehicle data, smart environment data, health applications on a smartphone, and Internet of Things in Ambient Assisted Living assure a promising future for healthcare. Medical image processing in radiology, sonography, MRI, laboratory methods, and automatic processing of skin conditions show robustness and precision.

Clinical decision support, robotic surgery support, and clinical workflow organization are benefiting from Big Data analytics methods. However, there is no consensus about Big Data's solely positive nature and its analysis.

#### 2.2.2 Data-driven and hypothesis-driven approach

Besides the problems with clinical Big Data listed above, there are serious concerns about the Big Data character itself [17]. There is a fundamental difference between traditional clinical data and big clinical data [18]. Traditional data is well prepared, structured, human-readable, and controlled, presented in relatively small sets, and easy to analyze with proven methods. Big data lacks a clear structure. It is often collected and handled without usual human supervision. Sometimes, the data remains unclear for users, even after classification and clusterization. Some researchers strongly believe that big data is more suitable for data-driven hypothesis creation. However, the proof of the hypothesis, especially in clinical research, still requires standard statistical methods and a traditional hypothesis-driven approach [19]. Multidimensionality incorporates large data sets, which are often not the case for clinical data. Sceptical voices reject even the possibility itself of Big Data replacing traditional statistics and metrics [20]. Traditional clinical biometrics and biomarkers leave limited options for completely unsupervised methods of data handling.

### 3. Potential solutions

A number of proposals have been made to tackle these issues. One approach is straightforward and oriented toward narrowing the gap between the human and machine sides. The other is called Explainable AI and deals with issues of a black box. An emerging field of large language models (LLM) can also be considered a prospective solution, at least partially. These options are discussed below.

### 3.1 Closing the gap

The difference between the human-managed data approach and automated big data collection and analysis can potentially be lessened if not entirely overcome. It can be achieved by making machines able to accept natural language information, by training human side operators in technical input-output specifics or by complex incremental approaches from both sides.

### 3.1.1 Machine side

Berners-Lee formulated one of the basic methods as a semantic web concept [21]. The gap between humans and machines can be closed from the machine side when all data becomes machine-readable. Then, only problems of data presentation and human interface will remain for the full solution. The concept still needs to be achieved.

### 3.1.2 Middle way

More mundane measures are needed to bring both sides closer. Up-to-date IT education, specifically for healthcare specialists, intuitive UIs, wide data standardization, and gradual, disruption-free technology implementation are practical steps toward workable solutions. CQL, the Clinical Quality Language, is human-machine readable and works in both domains [22]. A complex approach, when different sources of data are confluent, leaves a possibility for filtering all clinical data to subdivide it into small data sets for robustness and more straightforward operation.

### 3.1.3 Human side

The other data approach is patient oriented. The so-called 4P concept shows the way for future development. Personalized, predictive, preventative, participatory healthcare requires all types of data to be equally machine- and human-readable and analyzable.

The more mundane way to narrow the machine-human gap is to address the issue of practical usability habits. The prior video gaming experience is beneficial for operators in a high-fidelity virtual reality simulated robotic surgery [23]. Similar results are shown in real-day laparoscopic surgery by warming up through the video game prior to it [24]. Explainable AI can be related to the human-side solution. However, it is a matter of balance to put this topic into a separate part to discuss it in more detail.

### 3.2 Explainable AI

Big Data analysis with AI differentiates from regular data analysis with classical statistical tools. Unsupervised AI/ML neural networks, such as Deep Learning (DL), can be pretty efficient if judged by results but difficult for humans to comprehend. Black box AI poses a problem for high-risk areas of human activity. Explainable AI (XAI) is one way to solve it [25].

Often, ideal DL is envisaged as a tool that accepts raw data and produces high results with ready interpretation. There are methods to raise the accuracy of DL by tuning the neural network architecture and applying the algorithms with non-transparent machine-driven feature detection and processing. These features are usually hard to categorize for human readability, while the white box is readable and often connected to "real" or physical categories. At the same time, data processing in a black box can be highly effective and fast for large datasets. White box relies on supervised learning, which includes labelled data and supervised feature engineering. Semi-supervised learning is a combination of black-and-white boxes and might utilise partial feature engineering. Complex methods with ensemble meta-algorithms, such as boosting or bagging, belong to the category of white boxes. For example, Random Forest (RF) are often called "self-supervised". These methods are more explainable than the black boxes. Black box models, white box intrinsic or explainable by design models and grey box combinations have their strengths and weaknesses. For a description of structured characteristics, please see Table 1. The table demonstrates a negative relationship between performance and transparency in terms of human readability.

*Table 1: Comparative characteristics of black, grey and white box models*

| Type | Performance | Machine readability | Transparency | Human readability |
|---|---|---|---|---|
| Black box | ++++ | ++++ | + | + |
| Grey box | +++/++++ | +++/++++ | ++/+++ | ++/+++ |
| Explained White box | +++/++++ | +++/++++ | ++/+++ | ++/+++ |
| Intrinsic White box | ++ | ++ | ++++ | ++++ |

### 3.2.1 Black box

The opacity of AI models starts from the data collection stage. While limited data can be collected and interpreted by humans, big data, especially automatically collected or mixed, is less susceptible to human comprehension or analysis. The neural network was inspired by the formal model of the minimal neural equivalent of human intelligence or neurons. While the artificial "neurons" are pretty simple compared to biological specialised cells, the nature of NN "intelligence" is not much better understood than natural intelligence. Low transparency, in some way, is a result of the high capacity and robustness of the Deep Learning environment, where explainability is not required [26]. The limited control can be only applied in the stage of data collection, model training or output data interpretation. Multidimensional data analysis by unsupervised models is often more precise compared to more transparent ones. Nevertheless, higher precision is traded for obfuscation of analytical clarity for humans. Black box models can be explainable only post hoc, which is different from white box models (see Figure 3 for reference).

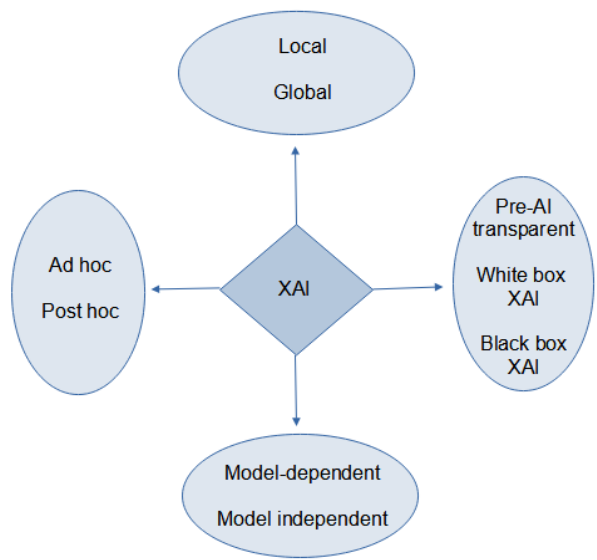

*Figure 4: Main categories of Explainable AI*

Main models with black box properties are based on neural networks, such as Convolutional Neural Networks (CNN), Recurrent Neural Networks (RNN), Directed Acyclic Graph Neural Networks (DAG-NN), and other types and combinations. More explainable, not black box, models include Generalized Additive Models (GAMs), random forest busting, bagging models, k-Nearest Neighbour (k-NN), and Support Vector Machines (SVM). The explainability of the last ones can be increased by applying the dimensionality reduction techniques where Principal Component Analysis (PCA) is one example of such types of approach. For schematic information, see Figure 4. The dependence between performance and transparency is formal and does not need to be strictly linear.

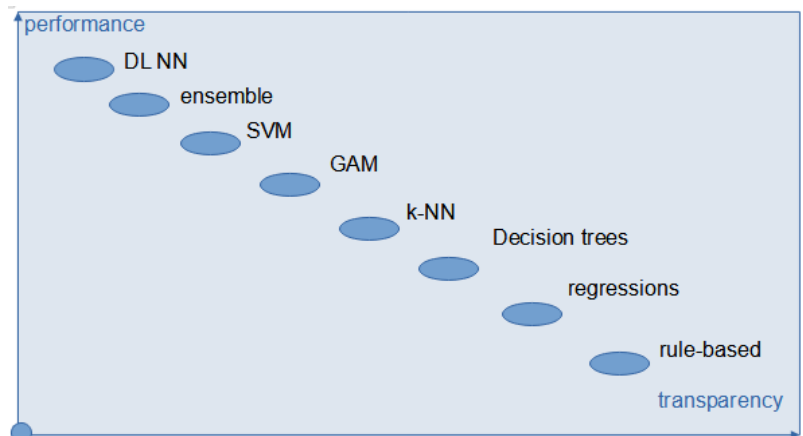

*Figure 5*: *AI/ML methods ranged by performance and transparency*

### 3.2.2 Ad hoc and post hoc

In most transparent cases, data is interpretable before AI/ML application. A human-interpretable, transparent by-design statistical AI model is called a white box. It is not a direct equivalent to supervised or semi-supervised AI/ML. White box frequently utilizes monotonic linear regressions (Linear Regression, Logistic Regression, LASSO), rule-based methods, and decision trees. Some of them perform only regression or classification, while others can do both. Probabilistic techniques, such as Naive Bayes, Markov models (can be seen as primary-level Dynamic Bayesian Networks) and other Bayesian networks are readable and often used for explainability.
The XAI approach can be subdivided into several main types [27] (Also see Figure 3). It is divided into local and global according to its explainability at the level of the neuron, layer, or whole system. According to the order of explainable model production, white boxes are ad hoc or intrinsic [26], while black boxes are potentially post hoc interpretable. The first ones are often designed to be interpretable (explainable by design), while the post hoc are produced in the process.
The post hoc [28] techniques of XAI include Local Interpretable Model-agnostic Explanations (LIMEs), which can be applied to any black box, as well as Anchor, another model-independent method [27]; Shapley Additive Explanations (SHAPs), sometimes applied with GAMs [29]; Layer-wise Relevance Propagation (LRP); Deep Taylor Decomposition (DTD); Explainable Graph Neural Networks (XGNN); Prediction Difference Analysis (PDA) and some others.

### 3.2.3 Grey box

The grey box models are usually a combination of black box and white box models. The main goal is to integrate the best characteristics of the black box and the white box, such as high performance and high transparency, respectively [26]. A grey box is suitable in situations where balance is needed between precision, confidentiality and openness. Proprietary algorithm performance can be united with model partial explainability without disclosure of private details. The grey box is a wide description of the model's spectrum with different levels and types of explainability, productivity and ability to focus these characteristics on specific areas of narrow interest.

## 3.3 Large Language Models

Large Language Models (LLMs) are recent popular objects of fascination for researchers and the general public. LLMs are extensive AI-powered chatbot versions that are able to communicate with users via natural language texts and generate prompted written content on demand. LLMs utilise various NLP techniques and generative AI. The most well-known LLMs, such as ChatGPT, BingGPT, Bart, Chinchilla, LLaMA, and Gopher, are primarily universal. There is also a high demand for specialized domain-focused LLMs.

### 3.3.1 Current successful applications

The first experiments in this area with ChatGPT [30] were promising when LLM was able to successfully pass the US medical qualification exams (USMLE) and master the Multiple-Choice Question Answering (MCQA) dataset for medical exams. A comparative study on the vascular surgery consultation scenarios was successfully done for GPT 3.5, GPT 4, Bard, and Falcon and compared to the panel of five field specialists [31]. Positive results are shown in the field of biomedical science and public health. Table 2 provides some examples of tests using LLMs applicable to healthcare.

*Table 2: LLMs in medicine and healthcare*

| Authors | Domain | Approach | Results and Conclusion |
|---|---|---|---|
| [31] | Consultation in vascular surgery emergencies | Consultation with Five attending surgeons and four LLMs | GPT-4 demonstrated 100% accuracy in emergency; further recommendations were accurate in 68%; 5% of responses were highly likely to cause clinically significant harm |
| [35] | Detecting Alzheimer's Disease-related signs and symptoms from electronic health records (EHRs) | Created taxonomy progression of AD signs and symptoms was trained using synthetic data generated by LLMs from EHRs. | The trained model demonstrated acceptable quality |
| [37] | Chatbot Systems in Chronic Cardiovascular Disease Self-Management | LLM-based chatbot system that supported chronic patients with hypertension | The system has potential to motivate patients to monitor their blood pressure and adhere to prescriptions |
| [38] | HealAI: A Healthcare LLM for Effective Medical Documentation | Optimized LLM for specific medical tasks | Optimized LLM outperformed GPT-4, maintaining precision, reducing biases, preventing hallucinations, and enhancing note-writing style |
| [39] | LLMs for Healthcare Data Augmentation | Privacy-aware data augmentation approach for LLM-based patient-trial matching (LLM-PTM) | Performance improved by 7.32% on average; the generalizability of new data improved by 12.12% |
| [40] | Evaluating LLM - Generated Multimodal | LLM-generated medical diagnosis with multimodal multiple- | GPT-4-Vision-Preview scored approximately 84% of correct diagnoses |

| | Diagnosis from Medical Images and symptom analysis | choice questions(MCQs) consisting of images and text for multiple pathologies | |
|---|---|---|---|
| [41] | Implications of LLMs for dental medicine | LLMs for administrative tasks | LLMs could improve administrative work via clinical decision support, text summarization, efficient writing, and multilingual communication, and can be applied to dental telemedicine |
| [42] | Impact of ChatGPT-3 on Health Services as a Virtual Doctor | Potential benefits from consultation for healthcare services, including the Internet of Orthopaedic Things (IOT) | Potential to become a Virtual Doctor (DocGPT). |

### 3.3.2 Limitations

Still, it is too early to make assumptions about LLM performance in the healthcare sphere. The well-known problem of LLM hallucination [32] limits not only the reliability and trustworthiness of the generated results but also creates a relatively new problem of responsibility for LLM-generated mistakes, wrong advice and externally coherent but medically meaningless solutions. GPT-4 demonstrated high interpretative abilities but showed reasoning inconsistencies, while Falcon and LLaMA 2 reached significant accuracy yet with insufficient explanatory reasoning [33]. The potential issues connected to the usage of LLMs are listed in Table 3.

*Table 3: Limitations of LLMs*

| Type | Possible Harm or Issues |
|---|---|
| Hallucinations | Production of a false output that does not match the user's intent |
| Security | Leakage of people's private information |
| Bias | Contribution to the spread of misinformation |
| Consent | Can expose users to copyright infringement issues |
| Scaling | Difficulties with time- and resource-consuming for scaling and maintaining the LLMs |
| Deployment | Requires deep learning, transformer model, distributed software and hardware, and overall technical expertise |

### 3.3.3 Potential solutions

There are several ways to address these problems (see Figure 5). The reliability of LLM-generated information can be increased by updating training sets, prompt engineering, domain-specific LLM training, real-time database connection of LLM, automatic output correction, and some other instruments and techniques [34]. For example, improvement in Out-of-Distribution (OoD) detection raises LLM performance [35]. In addition, protecting LLM from adversarial OoD attacks reassures LLM performance stability. Another promising method is the model self-reflection function [36].

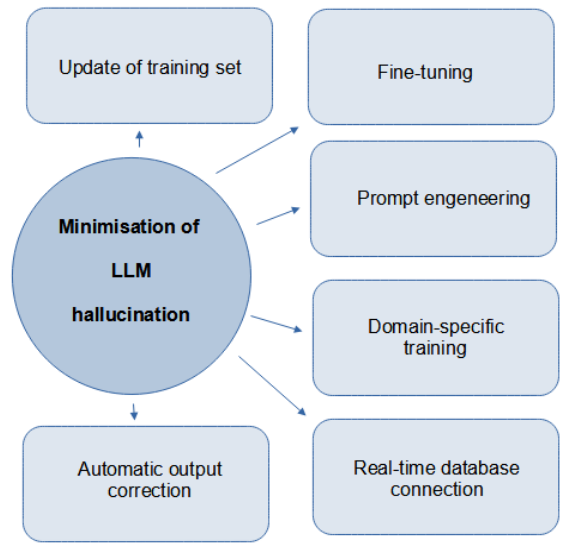

*Figure 6*: *Main methods of lowering LLM hallucination*

## 4. Conclusion

There are significant gaps between automatic and human data collection and analysis. These differences manifest themselves in the issues of Big Data and traditional statistical methods. Clinical data and its analysis are one of the testing grounds where questions about human control and responsibility reflect the complicated nature of these matters. Solutions are in the area of using more composite, heterogeneous human-collected and automatically collected data, mixed modes of analysis and dual readability for humans and machines. The tendency is complex: from one side, there is significant demand, based on technical capability, for automatically collected, analyzed and generated data. On the other side, there is no way to exclude human methods, which implies the necessity to enhance education and practical experience in the technical sphere not only for healthcare workers but for all members of modern society. Moreover, the implementation of modern LLMs might improve the diagnostic and treatment processes if the external reasoning agent is added to the models' architecture. It will increase the models' explainability and overall performance.

**Author Contributions:** Conceptualization, DJH and NJH; methodology, DJH and NJH; validation, DJH and NJH; formal analysis, DJH; investigation, DJH; resources, DJH; writing—original draft preparation, DJH; writing—review and editing, NJH; visualization, DJH and NJH. All authors have read and agreed to the published version of the manuscript.

**Funding:** No funds, grants, or other support was received.

**Competing Interest:** The authors declare that they have no conflict of interest.